\documentstyle [10pt,emulateapj]{article}

\begin{document}

\title{Low Mass Density Wide Field Far-IR/Submillimeter Telescope Systems }
\author{M. Dragovan\altaffilmark{1,2}} 
\altaffiltext{1}{Enrico Fermi Institute, University of Chicago, Chicago, IL 60637, mrk@oddjob.uchicago.edu}
\altaffiltext{2}{Jet Propulsion Laboratory, California Institute of Technology, Pasadena, CA  91125}


\begin{abstract}

Fundamentally new technology is described for constructing low areal mass density 
($\sim1$kg/m$^2),$ high precision ($< \sim10\mu$m RMS) reflectors scalable to 
large apertures ($\sim $10 to 20
meters) for use as the primary element of a telescope system.
A large reduction in mass is achieved by minimizing the mass of the 
reflective surface using a high reflectivity metallic membrane.
A wide field
diffraction limited telescope system can be constructed using the 
primary reflector in conjunction with
secondary and tertiary optics.

\end{abstract}



\section{INTRODUCTION}

Progress in observational astrophysics parallels the development
of telescope technology and the associated instrumentation. 
The problem of constructing telescopes using polished metal mirrors has a long history tracing
back to Gregory(1663), Newton(1672), and Cassegrain. The first
successful mirrors with silver reflecting surfaces on a glass substrate were constructed
in the late 1850's by von Steinheil and Foucault (King 1979).
Current state-of-the-art reflectors can trace their roots back to this 
technology. 

The function
of the substrate is to support 
the thin layer of high reflectivity material; the glass or metal substrate is
formable into a shape that has useful optical properties. 
In current state-of-the-art telescopes the mass of the substrate is
$10^3-10^6$ times the mass of the reflecting layer.
Clearly, new perspectives on telescope systems are necessary to 
reduce the cost and mass of the
primary element.

The technology described in this $letter$ achieves a significant
reduction in mass by minimizing the thickness of the substrate.
The telescope 
systems described use reflectors whose three dimensional shapes and
curvatures are formed by the bending or stretching of a membrane over an
appropriate boundary. The membrane is deformed by this process, with the
result that the surface assumes a shape that concentrates
electromagnetic radiation. If the field of view needs to be larger than can
be afforded by a single primary reflector, subsequent optics 
can correct the aberrations intrinsic to 
the primary. In either case, a diffraction limited system 
will result. By using suitable materials for the membrane and other
structures, systems with very low areal mass density 
($\sim 1$kg/m$^2)$ that are 
scalable to large
apertures ($\sim $10 to 20
meters) are constructable.

A number of designs for two and three mirror systems
have been developed, resulting in systems that have large focal
surfaces (Schroeder 1987). The systems
described are generally on-axis, where the secondary and tertiary optics
obstruct the primary reflector. Scattering and diffraction of the incident
electromagnetic radiation by the secondary optics and its support structure
reduces the performance of the overall system. This is particularly
problematic for observations of low-contrast objects, or in communications
systems where cross-talk between nearby antennas is undesirable.

The solution is to use an unobstructed, off-axis design. Unfortunately, the field-of-view
of such a system is limited unless steps are taken to control the new set of off-axis
aberrations. A solution using confocal conic reflectors was devised by Dragone (1982). 
Other designs, such as the aplanatic Gregorian or the Schwarzschild (1905; Claydon 1975)
solution have lower distortion and provide a wider field-of-view for off-axis 
systems at the expense of greater complexity of the surface shapes.

A wide
field-of-view off-axis
three element design is discussed, in which the secondary forms an image of
the primary on a third reflector. The combination of the secondary and
tertiary gives a wide field and also corrects for aberrations and defects
intrinsic to the primary reflector.

\section{AREAL DENSITY}

Current technology millimetric telescopes have densities of order 10kg/m$^2,$
a factor of $\sim 10^3$ between the mass of the reflecting layer and that of
the support structure. For optical telescopes the situation is much worse
where the current state-of-the art has density of order 150
kg/m$^2$, the supporting substrate $\sim 10^6$ times 
more massive than
the reflecting layer.

The areal density of the reflecting layer is given by
$\sigma _r=\rho_r t $
with $t$ the thickness of the reflecting layer, and $\rho_r $ the density. The
thickness of the reflecting layer of a high electrical conductivity metallic
film can be determined, to good approximation for a specific reflecting
material, by considering the skin depth
\[
\delta =1 / {\sqrt{\pi c \mu \sigma _e/\lambda}},
\]
where $\sigma _e\ $is the conductivity of the reflecting surface, $\lambda $
is the wavelength, and $\mu =400 \pi$nH/m. 
With a thickness of $t=7\delta$, the surface is opaque and the wave is 
reflected with low loss. 
For a very good conductor such as 
copper $\sigma _e=5.7\times 10^7 ( \Omega$m$)^{-1}$. 
In the case of optical light ($\lambda =0.5\mu$m) the film only has to
be $\sim 50$nm thick to reflect the incident light; for
millimeterwaves ($\lambda =1000\mu$m) a $\sim 1\mu$m thickness is required. 
This gives an areal density $\sigma _r\sim 8\times 10^{-3}$kg/m$^2$ in distinct contrast to
the areal density of the substrate material, which can be many orders of
magnitude greater.

By examining existing telescopes one finds that the areal mass density of the
supporting substrate (generally some form of glass) is
$\sigma \propto d^{0.5},$ where $d$ is the aperture diameter.
This is independent of the technology used, or the epoch when the telescope
was constructed.
In comparison, the areal density of a membrane reflector system scales
differently, and is straightforward to calculate. For the reflective
membrane 
\[
\sigma _r=\rho _rt_r.
\]
For the supporting boundary 
\[
\sigma _b={4\rho _bh(d) \Delta d} / d
\]
where $h(d)$ is the functional dependence of the boundary thickness with diameter,
and $\Delta d$ is the width of the boundary. The total
density is simply the sum
\[
\sigma =\sigma _r+\sigma _b=\rho _rt_r+{4\rho _bh(d) \Delta d} / d.
\]
It is instructive to note two cases, $h(d)=h$ (a constant height ring), and $%
h(d)=h_o(d/d_o)^{1/3}$ (a constant stiffness ring). In both cases the
areal density decreases with aperture. 
Only if the ring has $h(d)=h_o(d/d_o)^\alpha $ with $\alpha >1$ does $%
\sigma $ grow with $d,$%
\[
\sigma =\rho _rt_r+4\rho _{b\,}(h_o/d_o)\,(d/d_o)^{\alpha -1}\Delta d.
\]

This is in distinct contrast to the scaling relationship for existing 
telescopes, 
$\sigma \propto \,d^{0.5}.$ 
Thus, not only is a membrane reflector less massive to begin with,
but the areal density can actually {\it decrease} with larger apertures if
the ring and membrane are appropriately chosen.
Clearly, the areal density of a telescope system can be reduced
by orders of magnitude 
if the relativly massive supporting substrate can be minimized while
maintaining the desired reflective surface.

\section{Membrane Surfaces}

Deformable surfaces are naturally categorized by their Gaussian
curvature, an intrinsic property of any surface. 
All surfaces can be broadly categorized as 1) those that have zero Gaussian
curvature, and 2) all others. As is well known from differential
geometry the Gaussian curvature is given by $K = \kappa _1$ $\kappa _2$,
where $\kappa _1$ and $\kappa _2$ are the principal curvatures at a given
point on the surface. Membrane surfaces with both zero and non-zero Gaussian
curvature can be constructed and are considered in turn.

\subsection{Category 1: ${ K =0}${\bf \ }}

A surface with zero Gaussian curvature is either flat or has the shape of a
trough, so that one of the principal curvatures is always zero. Such a
surface can be formed by bending along only one axis. If the shape of the
surface in the curved direction is a parabola, then a line focus 
results for an incident plane wave. To produce a point focus, a 
system of two trough-shaped reflectors
properly oriented with respect to each other must be used. A perspective
view of such a system is presented in Fig 1. 
\centerline{{\vbox{\epsfxsize=7.0cm\epsfbox{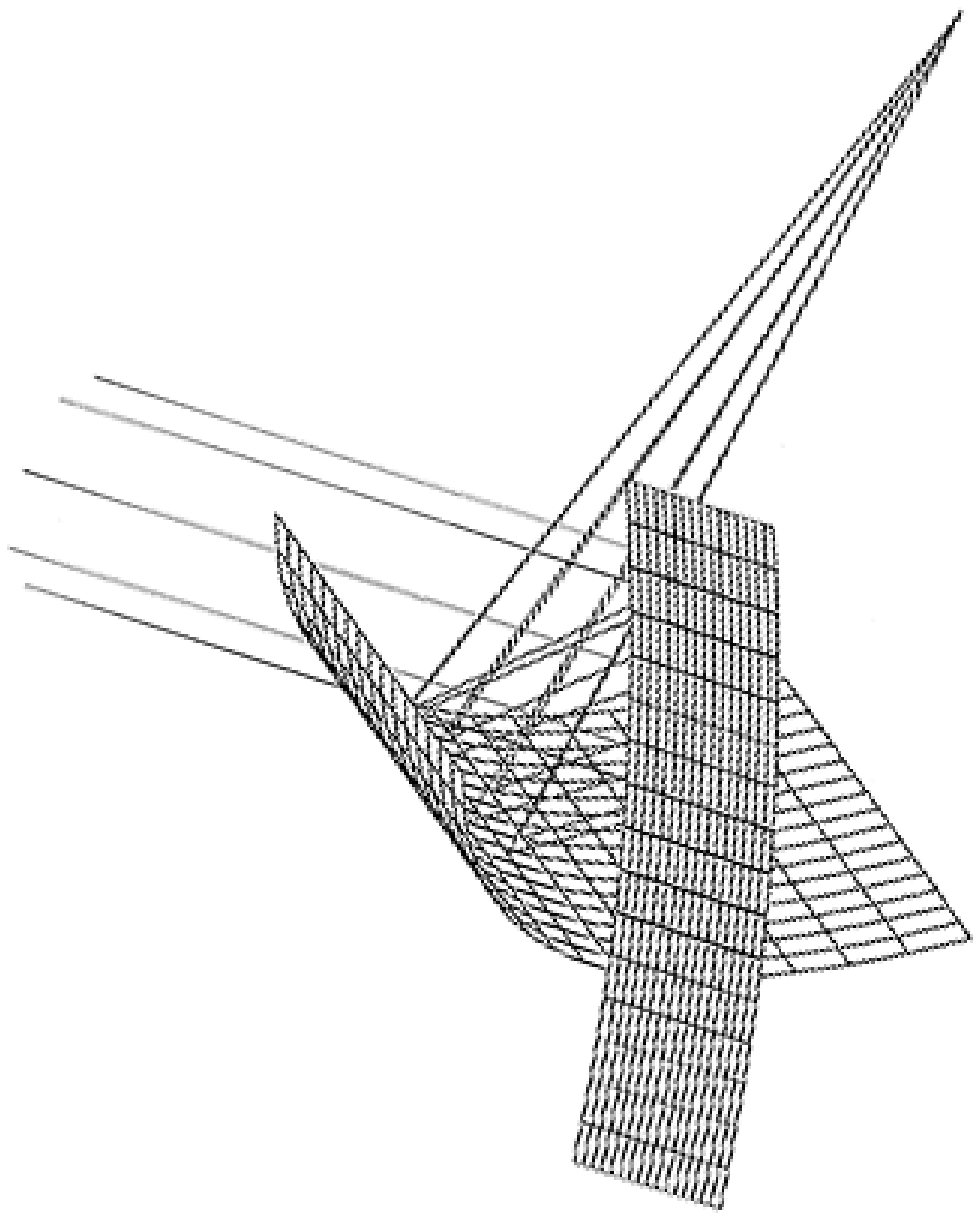}}}}
\figcaption{\footnotesize The layout of a two mirror reflector system
where the individual reflectors are parabolic cylinders. 
The orientation and curvatures of the individual reflectors
are chosen so that a point focus results for an
incident plane wave. The reflectors 
as illustrated are
greatly oversized to emphasize the curvatures of each reflective element.
It is clear by inspection
that the system is completely unobstructed. 
\label{fig1}}

In order for this system to focus and have a completely unobstructed aperture
the focal lengths of the two individual reflectors must be unequal. 
The aberrations of the system are identical to those of an off-axis paraboloid
with focal length $f_1$ in the direction which the first reflector focuses, 
and $f_2$ in the orthogonal direction, with the subscripts referring to the first or 
second reflector. For the specific system displayed in figure 1, the extent of the 
focal surface
is $30 \times 30$ diffraction limited pixels independent of wavelength.
The Airy disk is not circular, but has eccentricity 
\[
e=(1-(f_2/f_1)^2)^{1/2}.
\]

The parabolic-cylindrical surfaces are formed by tensioning a reflective
foil over a frame which has a parabolic contour along one axis and is rigid
enough to support the tensioning. 
The alignment of the two reflectors is critical to the performance of the
system. An arrangement of six adjustable rigid struts connecting the two 
reflectors completely constrains all degrees of freedom while allowing
the adjustment of the relative orientation of the two reflectors (Stewart 1965).

\subsection{Category 2: ${ K \neq 0}$}

Surfaces with non-zero Gaussian curvature can only be formed by stretching
or deforming a membrane along both axes. The shape and curvature the surface
assumes depends sensitively upon the boundary over which the membrane is
stretched, the pressure, and the mechanical material properties of the
membrane. If the boundary is circular an axisymetric reflector results.
However, the boundary need not be circular, nor planer; the only
requirement is that it is described by a space curve that closes upon
itself.
The surface constructed
has the useful optical property that it can concentrate 
electromagnetic radiation.

If the elastic limit of the material composing the membrane is exceeded, the
deformation is permanent and the pressure may be released resulting in
a self supporting reflector. If the elastic limit is not exceeded, a means
of maintaining tension in the membrane is necessary to hold the membrane in
the stretched state.
The tensioning of the membrane is accomplished by keeping a constant
pressure differential on the surface. This results in a membrane structure
that has uniform or continuous support over the entire
surface. 

\section{Plastically Deformed Membranes}

A pressurized membrane takes on a shape that minimizes the total energy of
the system consisting of the membrane, the gas and the structure rigidly
holding the membrane. There are three distinct shapes that can form,
depending upon how the surface tension, $\gamma ,$ in the membrane distributes
itself. Soap films, elastically deformed membranes, and plastically deformed
membranes are discussed.

\subsection{Modeling the membrane's shape}

Soap films take on shapes that minimize the energy of the film. Since the
energy of the film is proportional to the surface area, the question of
finding the minimal energy surface reduces to finding the surface of minimal
area satisfying the boundary conditions. The problem of finding minimal
surfaces is known as Plateau's problem, after the Belgian physicist 
who studied the problem using soap films
(Almgren 1969; Isenberg 1992).
An unpressurized soap film is a minimal surface and
has zero mean curvature $H=(\kappa_1+\kappa_2)/2 =0$. A pressurized soap film has 
non-zero constant mean curvature ($H \neq 0 $).

By choosing the appropriate curve for the boundary and pressurizing the film
so that it bulges out from the frame, convex or concave surfaces can be
formed. In the simplest incarnation the boundary can be chosen so that the
film is a segment of a torus, with the degenerate case being a circular
boundary which yields a spherical surface. This is both interesting and useful because a
segment of any surface with non-zero mean curvature can be approximated to
some degree by a toroidal segment (a surface that has two radii of curvature). An
off-axis segment of an ellipse or parabola can be fit quite well by a toroidal segment
(Cardona-Nunez et al. 1987).

The soap film is the special case of a membrane where the material 
can redistribute itself so that the tension $\gamma $
is everywhere constant. Next is a rigid membrane that remains elastic, but
has considerable deformation. Here $\gamma (r,\theta ),$ is {\it not}
constant as can easily be seen by noting that the at the boundary, where the
membrane is clamped, the tension is fixed; however, in the free radial
direction the tension can change (i.e. $\gamma (r,\theta )=\gamma (r)$). The
third case is that of plastic deformation. Again $\gamma (r,\theta )$ is not
constant, but when the pressure is released, the membrane {\it does not}
return to its original state. This is the situation of the membrane
reflectors. A variational analysis can be performed for the elastic and
plastic flow cases (Murphy 1987; Weil and Newmark 1955),
with the result that the shape of the membrane is well
approximated as a conic section with higher order polynomial deviations.

The model used to predict the shape of the deformed surface is that of
the soap bubble, where the metallic membrane has been sufficiently 
deformed so that this is
a good approximation. 
The governing equation for the surface $u(x,y)$ is
\[
{\bf \nabla }\cdot \left( \frac{{\bf \nabla }u}{\sqrt{1+|{\bf \nabla }u|^2}}%
\right) =p/2\gamma =\left( \frac 1{r_1}+\frac 1{r_2}\right),
\]
where $r_1$ and $r_2$ are the two radii of curvature of the surface, and $p$ 
is the pressure difference between the inside and outside of the membrane
 (Bateman 1932; Struik 1961). This non-linear equation 
may be solved numerically given the geometrical boundary, the pressure, 
and the surface tension.

\subsection{Fabrication and Characterization}

A rigid volume is capped with a thin, flat,
reflective, and stretchable material (the membrane). The volume is pressurized 
with a gas, similar to inflating a balloon or soap bubble. 
The resulting membrane surface is analytically represented by a conic section with
polynomial correction terms. 

The surface profile of a free standing plastically deformed membrane was measured
using a non-contacting laser displacement sensor. The data were fit to a parabola 
in order to evaluate the magnitude of the wavefront error that needs to be corrected.
The residuals of the fit are
plotted in fig 2.  
\centerline{{\vbox{\epsfxsize=7.0cm\epsfbox{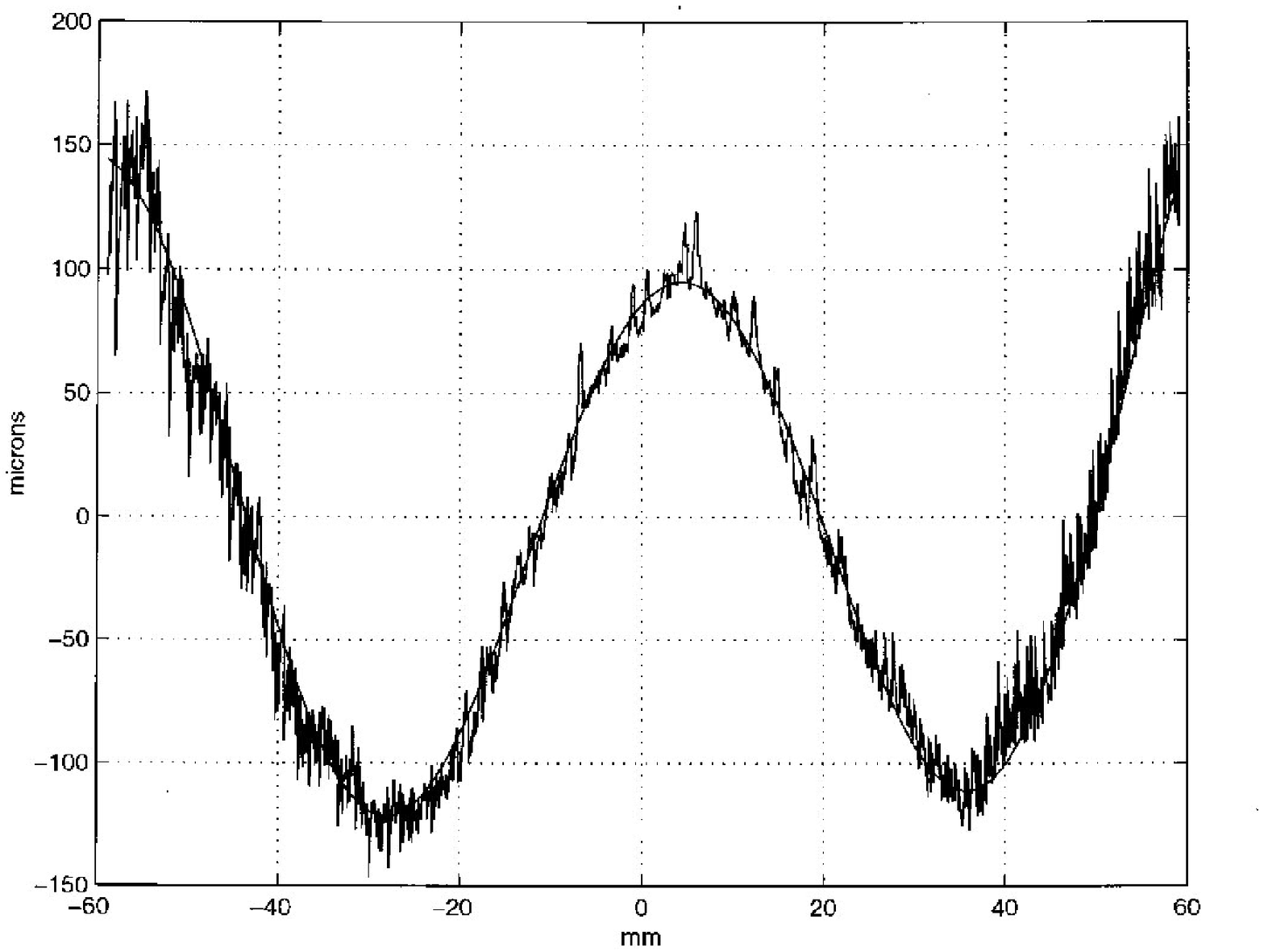}}}}
\figcaption{\footnotesize The residuals of the profile measurements of
the free-standing plastically deformed reflector. The residuals have a characteristic
polynomial shape. After fitting to a 8th order polynomial the surface roughness is
found to be 8 $\mu$m RMS.
\label{fig2}}

The membrane is a $50 \mu$m rolled stainless steel foil with 
fabricated surface finish of $\sim 8 \mu$m RMS. This is the same value 
measured after the surface measurements were fit to a parabolic + 8th order 
polynomial. The reflector as constructed is suitable for use in the 
far-infrared/submillimeter: the small scale surface roughness having a value
of $\sim\lambda /25$, and the global wavefront error of $\sim \pm 1 \lambda$ 
is correctable as outlined below.
Copper, silver, and nickel membranes have also been used and give similar 
results.

\section{Secondary and Tertiary Reflectors}

In the case of a parabolic primary it is well known that an elliptical 
(Gregorian) or hyperbolic (Cassegrain) secondary results in a telescope system
with an improved field-of-view. Even better performance can be achieved by 
adjusting the conic constants of the primary and secondary so that the Abbe
sine condition is minimally violated. This conic approximation is the
basis of the Ritchey-Chretien solution. Schwarzschild (1905) solved the problem
of finding the analytic form of the surfaces that exactly
satisfy the sine condition.
In the following sections algorithms are given that generate 
the shapes needed for corrective secondary and tertiary reflectors.

\subsection{One Reflector Correction}

A solution to the problem of correcting a spherical reflector with
a single secondary is given by Head (1957) and Geruni (1964). The results
are generalizable to non-spherical primary reflectors.
Since only the secondary is adjustable the field-of-view of this system is
smaller than that of either the classical Gregorian or the Ritchey-Chretion.

\subsection{Two Reflector Correction}

An aplanatic system is free of both astigmatism and comma, and
has the widest focal surface of any design.
An algorithm to generate aplanatic surfaces is described by 
Lundberg (1964).  This algorithm can be generalized to generate corrective secondary
and tertiary reflectors with a given primary reflector.
Two conditions on the ray paths through the system are (Figure 3): 
1) the Abbe sine condition 
\[
\frac h{sin(\theta)}=k=|x_2-x_1|+|x_2-x_3|+{((x_3-x_4)^2 + y_4^2)^{1/2} }; 
\]
2) constant path lengths for each ray from the entrance aperture to the focal point 
\[
l=k+|x_0-x_1|.
\]
The surfaces $M2(x)$ and $M3(x)$ are generated 
recursively by tracing 
rays through the system. The rays hit $M1$ and are deflected
towards $M2$ and $M3$; $M2$ is adjusted so that the path length condition is 
satisfied while $M3$ is adjusted so that the sine condition is meet.
A visualization is provided by considering individual rays as 
strings of constant length. Where they contact a reflective surface the 
angles must satisfy the reflection law. 
With this procedure $M2(x)$ and $M3(x)$ are generated. 

The best performance is obtained when tertiary M3 is 
located near the image of the primary produced by M2. This location is ideal
for correcting the shape imperfections of the primary reflector. This can be
seen by noting that a region on the primary is imaged to a region on the
tertiary via the secondary reflector. Thus, by distorting the tertiary
opposite to the shape imperfection of the primary a corrective system is
produced. This applies to all scales of spatial imperfections, with the large
scales being the easiest to correct.

\subsection{Tertiary Scanning }

A complementary use for the tertiary is to allow fast scanning or tracking
of an observation point by rotating the tertiary about an axis orthogonal 
to the optical axis. Since the tertiary is located at an image of the
primary, this is optically equivalent to rotating the primary. Rotation of the
tertiary effects a change in the observing direction without changing the
illumination of the primary reflector. The tertiary is
smaller than the primary, so the scanning or tracking is accomplished
with high performance.
\centerline{{\vbox{\epsfxsize=7.0cm\epsfbox{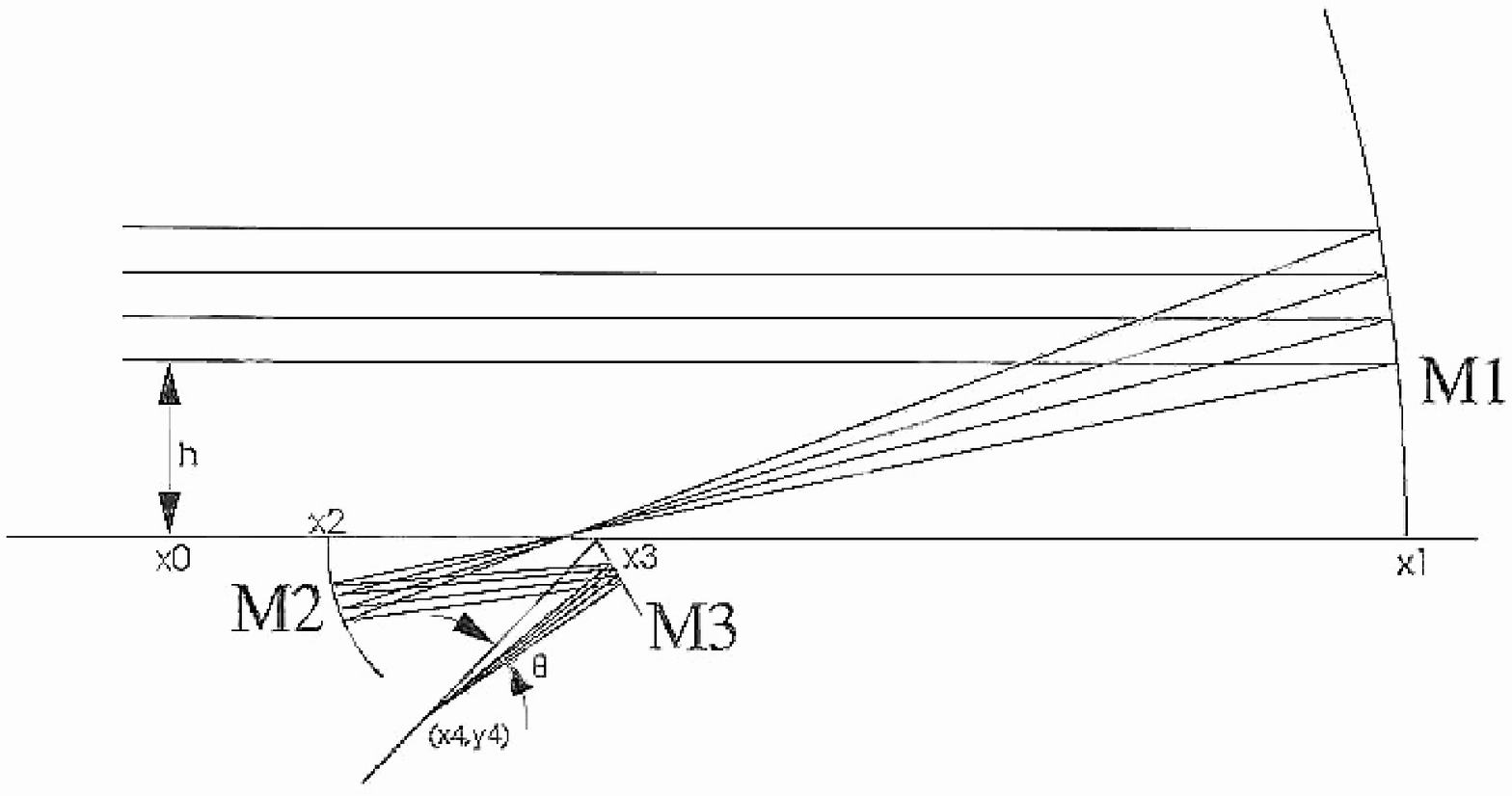}}}}
\figcaption{\footnotesize The 2 reflector corrector 
system (M2, M3) generated using a modified Lundberg algorithm. 
$M2(x)$ and $M3(x)$ are adjusted
so that the conditions of constant path length and the Abbe sine condition are meet.
In a complementary use, M3 may be 
rotated to effect scanning on the sky.
\label{fig3}}

\section{Summary and Conclusions}

This letter describes arrangements of reflectors that 
concentrate electromagnetic radiation with the primary reflecting surface 
formed by either stretching or tensioning a membrane over a suitable boundary.
The systems are constructed so that the resulting telescopes 
have very low areal mass density 
($\sim 1$kg/m$^2$) and are scalable to large apertures with diffraction limited performance. 

\acknowledgments
I wish to thank my CARA, JPL, NSF, and STACEE colleagues for their criticisms and suggestions. 
Former students 
B. Crone and R. Leheany assisted in the initial stages of this work;
G.H. Marion and A.K. George assisted with the construction and 
characterization of the plastically deformed reflector.
This work was supported by NSF, the McDonnell foundation, NASA, 
and the Fullahm award of the
Dudley observatory.


\begin{references}

\reference{} Almgren, F.J. 1969 {\it Plateau's Problem.} Benjamin.

\reference{} Bateman, H. 1932 {\it Partial Differential Equations of Mathematical Physics.}
Cambridge.

\reference{} Cardona-Nunez, O. et al. 1987 {\it Applied optics } {\bf 26}, 4832.

\reference{} Claydon, B. 1975 {\it Marconi Review } {\bf 1st quarter}, pp 14-43.

\reference{} Dragone, C. 1982 {\it IEEE Trans. Ant. Prop. }
{\bf AP-30}, 331.

\reference{} Geruni, P.M 1964 {\it Radiotechnica i Electronika } (Eng. trans.), {\bf 9}, 1.

\reference{} Gregory, J. 1663, {\it Optica Promota }.

\reference{} Head, A.K 1957 $Nature$, {\bf 179}, 692.

\reference{} Isenberg, C. 1992 {\it The Science of Soap Films and Soap Bubbles.} Dover.
\reference{} King, H.C. 1979 {\it The History of the Telescope.} Dover.

\reference{} Lundberg, R.K. {\it Mathematical Theory of Optics. } University of
California Press, 1964 (p208).

\reference{} Newton, I. 1672 {\it Phil. Trans.,} {\bf 7}, pp 4004-10.

\reference{} Murphy, L.M. 1987 {\it Journal of Solar Energy Engineering} {\bf 109, }111-120

\reference{} Schroeder, D.J. 1987 $Astronomical$ $Optics.$ Academic Press.

\reference{} Schwarzschild, K. 1905 {\it Astr. Mittheilungen Kognigl.} Sternwarte Gottingen, {\bf 10},1.

\reference{} Stewart, D. 1965 {\it Proc. Instn. Mech. Engrs. } {\bf 180}, 371.

\reference{} Struik, D.J. 1961 {\it Lectures on Classical Differential Geometry.}
Addison-Wesley.

\reference{} Weil, N.A. and Newmark, N.M. 1955 {\it Journal of Applied Mechanics,} Dec.,pp 533-538.

\end{references}
\end{document}